\documentclass[%
:reprint,
superscriptaddress,
preprint,
 amsmath,amssymb,
prb,
]{revtex4-2}

\usepackage[a4paper, paperwidth=17.2cm, paperheight=25cm, inner=1.5cm, top=1.5cm, outer=1cm, bottom=1cm, includefoot]{geometry}
\usepackage{bm}
\usepackage{graphicx}
\newcommand{\bx}{{\bm x}}
\newcommand{\kb}{k_{\rm B}}
\newcommand{\<}{\langle}
\renewcommand{\>}{\rangle}

\usepackage{color}
\definecolor{ascol}{rgb}{0.7,0, 0}

\begin{document}
	
\title{Bayesian learning of thermodynamic integration and numerical convergence for accurate phase diagrams}

\author{V. Ladygin}
\email{vladimir.ladygin@phystech.edu}

\affiliation{Moscow Institute of Physics and Technology, Institutskiy Pereulok 9, Dolgoprudny, Moscow Region 141700, Russia}
\affiliation{Skolkovo Institute of Science and Technology, Skolkovo Innovation Center, Building 3, Moscow 143026, Russia}
\affiliation{California Institute of Technology, 1200 E California Blvd, Pasadena, CA 91125, USA}

\author{I. Beniya}
\affiliation{Skolkovo Institute of Science and Technology, Skolkovo Innovation Center, Building 3, Moscow 143026, Russia}

\author{E. Makarov}
\affiliation{Skolkovo Institute of Science and Technology, Skolkovo Innovation Center, Building 3, Moscow 143026, Russia}

\author{A. Shapeev}
\email{a.shapeev@skoltech.ru}
\affiliation{Skolkovo Institute of Science and Technology, Skolkovo Innovation Center, Building 3, Moscow 143026, Russia}

\date{\today}

\begin{abstract}
Accurate phase diagram calculation from molecular dynamics requires systematic treatment and convergence of statistical averages.
In this work we propose a Gaussian process regression based framework for reconstructing the free energy functions using data of various origin.
Our framework allows for propagating statistical uncertainty from finite molecular dynamics trajectories to the phase diagram and automatically performing convergence with respect to simulation parameters.
Furthermore, our approach provides a way for automatic optimal sampling in the simulation parameter space based on Bayesian optimization approach.
We validate our methodology by constructing phase diagrams of two model systems, the Lennard-Jones and soft-core potential, and compare the results with the existing studies and our coexistence simulations.
Finally, we construct the phase diagram of lithium at temperatures above 300 K and pressures below 30 GPa from a machine-learning potential trained on ab initio data.
Our approach performs well when compared to coexistence simulations and experimental results. 
\end{abstract}

\maketitle
\section{Introduction}

Computational materials science is a rapidly evolving field enabling the calculation of materials properties that have traditionally been accessible mostly by experiment.
Phase diagram is one such aggregate property; it answers the following question: which phase of material will be stable under given conditions (temperature, pressure, composition)?
Phase diagrams are thus indispensable as guidance for materials synthesis.
Physically, a phase diagram can be thought of as a map of free energy of different phases of material; if we know the free energy function of different phases, then we can tell which phase, or a mixture of phases, will be stable under a given condition.


CALPHAD (standing for CALCulation of PHAse Diagrams) \cite{saunders1998calphad,spencer2008brief} is by far the most prominent approach to constructing phase diagrams in practice.
The core of CALPHAD is a classical fitting approach, using polynomial-like functions, to represent the Gibbs free energy functions of different phases.
The free energy is fitted mostly to the experimental data, and in this sense, the obtained phase diagram is an experimentally obtained materials property (at least when contrasted against modern ab initio-based materials modeling).
With the rise of ab initio materials modeling, the experimental data can be supplemented by quantum-mechanical data when the former is not available (e.g., when a phase cannot be experimentally realized at given conditions).

There are efforts to further advance the CALPHAD methodology by applying modern data analysis algorithms.
Early attempts are dedicated to the use of the Bayesian framework in combination with the classical CALPHAD approach \cite{konigsberger1995new,konigsberger1991improvement,stan2003bayesian}. In these works, the authors focus on a methodology for uncertainty prediction of model parameters of CALPHAD. More recent studies are aimed at overcoming the problem of database extension \cite{bocklund2019espei} with the use of first-principle data calculation. The further development of the approach \cite{bocklund2019espei} includes uncertainty estimation of predicted results based on uncertainty in model parameters via Monte Carlo Markov chains combined with the Bayesian inference \cite{otis2017high,paulson2019quantified}. Another problem related to a mixture of experimental and calculation results was approached recently in \cite{zomorodpoosh2020statistical} with k-fold cross-validation of the input datasets. 




From the side of computational materials science, state-of-the-art algorithms for the free energy computation are grouped together under the umbrella of thermodynamic integration.
The central idea of these algorithms is that one can easily compute derivatives of free energy from molecular dynamics (MD) simulations (or similar simulations for saving the Gibbs distribution), and hence the free energy can be obtained by integrating these data from a point where the free energy can be calculated exactly (e.g., at zero temperature) \cite{frenkel2001-md-book}.
Thermodynamic integration, in its essence, is a way to obtain computational thermodynamic data, and hence is not an algorithm that is competing with CALPHAD, but rather complementing it. 

Indeed, there are many works that use both ideas: obtaining computational thermodynamic data and fitting it with simple functions of thermodynamic parameters for a number of systems.
Examples of such works are studies of the Lennard-Jones system \cite{mastny2007melting} where the authors examine the dependence of melting temperature with respect to the size of the simulation cell. In \cite{sjostrom2016multiphase}, the reconstruction of the aluminum phase diagram via DFT data is presented. In \cite{kruglov2019phase} authors reconstruct the P-T phase diagram of Uranium with a combination of USPEX algorithm \cite{glass2006uspex} and standard thermodynamic integration technique for the free energy calculation.
In the work \cite{kruglov2019phase} authors focus on phase stability calculation under finite temperature conditions in specific points.
In \cite{reinhardt2021quantum} a detailed phase diagram of water is explored.

In this work, we propose a new, Gaussian process-based methodology to reconstructing the phase diagram based on thermodynamic data.
Although we focus on non-experimental data (computed from first principles or with empirical potentials) in the work, the methodology itself should be applicable to the experimental data as well.
The essence of our methodology can simply be described as treating thermodynamic data coming from any source (zero-temperature limit or MD averages or coexistence simulations) as training data for a Gaussian process.
Including numerical parameters (such as the number of atoms or the cutoff radius of the potentials) into the feature vector enables an automatic analysis of convergence and, moreover, taking the limit as numerical parameters go to infinity.
Furthermore, the predictive variance of the Gaussian process naturally allows us to estimate the error of our prediction, including the statistical error originating from the finite MD trajectories, interpolation error arising from the finite number of conditions (temperature/pressure) at which simulations are run, and the error of extrapolation with respect to numerical parameters.
Finally, Gaussian processes allow for an automatic assessment of the ratio by which the error in the quantity of interest (such as the melting temperature) can be reduced by running a simulation at given parameters, leading to autonomous algorithms of sampling the phase diagram points.


The presented methodology is applied to two model systems: soft-core and Lennard-Jones potentials as a part of the validation procedure.
We compare our calculations with the work \cite{morris2002melting} on the soft-core potential in which the authors estimated the melting curves in a wide temperature-pressure range.
The Lennard-Jones system's properties, including the dependence on the simulation parameters, were investigated in various works.
The points of interest on the Lennard-Jones phase diagram are the critical point where gas and liquid become indistinguishable and the triple point where all the three phases---gas, liquid, and solid---coexist. The location of the critical point strongly depends on the size of the simulation cell. In literature, there are two main approaches to study this dependence. The first one is associated with the projections of a simulation cell with periodic boundary conditions into the surface of a 4-dimensional sphere \cite{caillol1998critical}. Another approach relies on the use of cutoff radius bounded to the half-size of the simulation cell, and corresponding long range correction \cite{perez2006critical}. The critical point temperature estimations are not in agreement with each other with the given narrow confidence intervals. This fact shows that the approach for systematic phase transition calculation and uncertainty estimation is in demand in computational materials science.
For validation of our results with respect to triple and melting points calculation of the Lennard-Jones system, we have chosen the work \cite{mastny2007melting}.

Finally, we apply our approach to a physical system, lithium, chosen because it undergoes various phase transitions under pressure. The CALPHAD approach for this material is based on experimental data from \cite{luedemann1968melting, lazicki2010high, schaeffer2012high}. These works are attributed to fcc-liquid-bcc phase transitions. In \cite{guillaume2011cold} the phase diagram of lithium is examined in a wide temperature and pressure range from first principles complementing experiment. Recently, in \cite{dorrell2020pressure} authors compute the former transition lines with the existing classical potentials.

The paper is organized as follows. In Section \ref{sec:theory} we present the theoretical aspect of the methodology and some implementation details.
In particular, in Section \ref{sec:free_en} we give the details of free energy calculations and in Section \ref{sec:GP} we introduce our Gaussian process regression approach.
In Section \ref{sec:res} we show the results of the application of our approach to the phase diagram calculation of the model systems and lithium.
Concluding remarks are given in Section \ref{conc}.

\section{Theory}\label{sec:theory}

In the current section, we will obtain the relations between free energy derivatives and statistical averages as will be used by Gaussian processes.
In what follows, we distinguish the extensive and intensive quantities in the notation: the former will have a hat accent: $\hat{E}$, $\hat{V}$.
The corresponding intensive, per-atom quantities are $E = \hat{E}/N$, $V = \hat{V}/N$, where $N$ will denote the number of atoms.

\subsection{Free energy}\label{sec:free_en}

Let $\bx$ be a configuration with $N$ atoms enclosed in a volume $\hat{V}$ (we will interchangeably use $\hat{V}$ for the actual region in space and its measure) with periodic boundary conditions.
Let $\hat{E}(\bx)$ be the potential energy of the interatomic interaction.
We assume that the units for the temperature $T$ are the same as for the energy $\hat{E}$; in other words, our Boltzmann constant is $\kb = 1$.



We define the free energy by
\begin{equation}\label{eq:free_energy}
-T \log \int_{\hat{V}^N} \exp(-\hat{E}(\bx)/T) {\rm d}\bx = \hat{F}_{\rm ref} -T \hat{S},
\end{equation}
where $\hat{F}_{\rm ref}$ will be explicitly assigned later (differently for each phase), and we will call $\hat{S}$ the entropy.
We will rely on comparing the absolute free energies of different phases (as opposed to the free energy difference between phases); therefore, it is important to choose $\hat{F}_{\rm ref}$ and $\hat{S}$ consistently across phases.

For solid we choose 
\begin{equation} \label{eq:F_ref_sol}
F^{({\rm s})}_{\rm ref} := E_0 + T \big( -\log(NV) + 1 -{\textstyle\frac{3}{2}} \log (2\pi T) \big),
\end{equation}
where $E_0=E_0(V)$ is the potential energy at zero temperature for the given volume.
Here and in what follows the superscript ${\rm (s)}$ denotes the solid phase. 
Choosing $F^{({\rm s})}_{\rm ref}$ in the form \eqref{eq:F_ref_sol} is motivated by the fact that in this case the entropy admits a simple low-temperature expansion
\begin{equation} \label{eq:entropy_sol}
S^{({\rm s})}
=
- {\textstyle\frac{1}{2 N}} \log \det\hat{H} - \log(V)
+ O(T),
\end{equation}
where $\hat{H}$ is the Hessian of the energy $\hat{E}$ at the equilibrium configuration $\bx_0$.
For the derivation of \eqref{eq:F_ref_sol} and \eqref{eq:entropy_sol}, refer to Appendix \ref{sec:solid-ref}. 
A well-defined zero-temperature limit of $S^{({\rm s})}$ such as \eqref{eq:entropy_sol} is important for reconstructing the free energy with a Gaussian process, as molecular dynamics can generate the data on derivatives of $S$ and thus allow, without \eqref{eq:entropy_sol}, one to reconstruct $S^{({\rm s})}=S^{({\rm s})}(T,V)$ only up to an additive constant.

Indeed, an NVT-thermostatted molecular dynamics produces the averages of the form
\[
\< f \> := \frac{\int_{\hat{V}^N} f(\bx) \exp(-\hat{E}(\bx)/T) {\rm d}\bx}{\int_{\hat{V}^N} \exp(- \hat{E}(\bx)/T) {\rm d}\bx}.
\]
One can then find that
\begin{align}\label{eq:e_dir_sol}
	\frac{{\rm \partial}S^{({\rm s})}}{{\rm \partial}T} 
	&=
	T^{-2} \<E-E_0\> - {\textstyle\frac{3}{2}} T^{-1},
	\quad\text{and}\\ \label{eq:p_dir_sol}
	\frac{{\rm \partial}S^{({\rm s})}}{{\rm \partial}V}
	&=
T^{-1} \< P - P_0 \>
	,
\end{align}
where $P_0 := -\frac{\partial E_0}{\partial V}$ is the pressure at zero temperature. Here and in what follows by $P$ we denote the virial part of the pressure.
The virial pressure is, in fact, easier to compute from molecular dynamics.
The derivation of \eqref{eq:e_dir_sol} and \eqref{eq:p_dir_sol} is given in Appendix \ref{sec:solid-der}.

Also, we consider the liquid and gas phases.
A single free energy curve can describe these phases because they are indistinguishable at temperatures above the critical one.
Hence, we will denote the corresponding phase by superscript ${\rm (f)}$ and refer to it as the fluid phase.
For the fluid, we simply choose ideal gas as a reference,
\begin{equation} \label{eq:F_ref_liq}
	F^{({\rm f})}_{\rm ref} := - T \log (N V),
\end{equation}
so that 
\begin{equation}
\lim_{T\to\infty} S^{({\rm f})}
= 
\lim_{V\to\infty} S^{({\rm f})}
= 0.
\label{eq:liq_lim}
\end{equation}
This equality is a consequence of our definition of free energy in \eqref{eq:free_energy}. In the limit of $T \rightarrow \infty$, the exponent in the integral is approaching one. Hence, the integral itself is equal to $N V$. In the same manner, when $V \rightarrow \infty$, the interaction between particles is negligible ($\hat{E}(\bx)\rightarrow 0$) and the integral also approaches $NV$.

The derivatives of $S^{({\rm f})}$ are thus
\begin{align}\label{eq:p_dir_liq}
	\frac{{\rm \partial}S^{({\rm f})}}{{\rm \partial}V}
	&=
	T^{-1} \< P\>,
\quad\text{and}\\ \label{eq:e_dir_liq}
	\frac{{\rm \partial}S^{({\rm f})}}{{\rm \partial}T} 
	&=
	T^{-2} \<E\>.
\end{align}
The derivation of \eqref{eq:p_dir_liq} and \eqref{eq:e_dir_liq} is very similar to the corresponding formulas for the solid, hence we omit such a derivation.

The harmonic \eqref{eq:F_ref_sol} and ideal gas \eqref{eq:F_ref_liq} limits are not always applicable---for instance, there are systems with solid phases being dynamically unstable at low temperature. For such systems, the limit \eqref{eq:F_ref_sol} is irrelevant.
In such cases, we determine the additive constant of the free energy through fitting to the melting (or, more generally, coexistence) point of a phase. We find melting point at pressure $P$ by solving the system of equations
\begin{equation}\label{eq:trans_Y}
	\left\{
\begin{array}{l}\displaystyle
\frac{ \partial S^{\rm(f)}}{\rm  \partial V^{\rm (f)}} = \frac {P} T + \frac 1 T \frac{ \partial F^{\rm(f)}_{\rm ref}}{ \partial V^{\rm (f)}} \\
\\ \displaystyle
\frac{ \partial  S^{\rm(s)}}{ \partial V^{\rm (s)}} = \frac {P} T + \frac 1 T\frac{\partial F^{\rm(s)}_{\rm ref}}{  \partial V^{\rm (s)}} \\
\\ \displaystyle
S^{\rm(f)} - S^{\rm(s)} = \frac {F^{\rm(f)}_{\rm ref} -  F^{\rm(s)}_{\rm ref}} T + \frac{P \big(V^{\rm (f)} - V^{\rm (s)}\big)}{T},
\end{array}
\right.
\end{equation}
with respect to the temperature $T$ and specific volumes of solid and fluid, $V^{\rm (f)}$ and $V^{\rm (s)}$.

%
%
%

%

\subsection{Gaussian process regression \label{sec:GP}}

The derivatives of the entropy from an NVT molecular dynamics (MD) cannot be obtained without some noise arising from averaging over a finite trajectory.
Due to randomness in the initial state or in the thermostat, such a trajectory is random.
Thus, the free energy that we reconstruct from the MD data is also random, but hopefully, it has a narrow distribution around the true free energy.
The effect of a thermostat---let us consider a Langevin thermostat, for instance---consists of making a large number of small perturbations to the trajectory \cite{allen2017computer-book}.
Thanks to the central limit theorem, it is hence reasonable to assume that averages over such a trajectory are distributed according to the Gaussian distribution.
This assumption brings us to the Gaussian process framework.

In the Gaussian process framework, we assume that the data, and the reconstructed free energy, are distributed according to a multivariate Gaussian distribution.
We assume zero mean---any prior information about a nonzero mean is already accounted for in $F^{\rm ref}$.
Further, we assume that the values of the free energy at different points $(V_1,T_1)$ and $(V_2,T_2)$ are correlated with covariance $\operatorname{Cov}(S(V_1,T_1), S(V_2,T_2)) = k\big((V_1,T_1),(V_2,T_2)\big)$.
Such a distribution of functions $S(V,T)$ is called the \emph{Gaussian process} (GP) and $k$ is called \emph{the kernel}.
An simple example of the kernel is 
\begin{equation}\label{eq:kernel-example}
k ( (V_1,T_1), (V_1, T_1) ) \sim \exp\left(- \frac {(T_1 - T_2)^2} {2 \theta_T^2}\right)  \exp \left(- \frac {\left( V_1 - V_2 \right)^2}  {2 \theta_{V}^2 } \right).
\end{equation}

A property of Gaussian processes that will be very helpful in our application is that any linear functional of the Gaussian process is also Gaussian-distributed.
For example, the derivative with respect to volume (as, e.g., in \eqref{eq:p_dir_liq}) at $(V_1, T_1)$ is correlated with $S(V_2, T_2)$ as given by the following formula:
\[
\operatorname{Cov}\bigg(
\frac{\partial S}{\partial V_1} (V_1,T_1), S(V_2,T_2)
\bigg)
=
\frac{\partial}{\partial V_1} k\big( (V_1,T_1), (V_2,T_2) \big).
\]

This allows us to make predictions based on data.
In the most general case, each data point is a linear functional $X$ on $S$, for example $\<S | X_1 \> = S(V,T)$, $\<S | X_2 \> = \frac{\partial}{\partial V}S(V,T)$, etc.
The data is usually given with noise, hence the input data to our Gaussian process is of the form
$	(X_1, Y_1, \Delta Y_1),
	(X_2, Y_2, \Delta Y_2), \ldots,
$
which means that $\<S | X_i\>$ is measured (e.g., from molecular dynamics) as $Y_i\pm \Delta Y_i$, or to be precise, $\<S | X_i\>$ is distributed according to the normal distribution
\[
\<S | X_i\> \sim \mathcal{N}\big(Y_i, (\Delta Y_i)^2 \big).
\]
We denote $\operatorname{Cov}\big(\<S | X_1\>,\<S | X_2\>\big) = k(X_1,X_2)$ extending the definition for the kernel.
We assume that the uncertainties $\Delta Y_i$ are all statistically independent from each other.

Suppose we want to make a prediction of $Y_* = \<S | X_*\>$; for simplicity one can think of $Y_* = F(V_*,T_*)$.
The Gaussian process framework is a particular case of the Bayesian framework in which the prediction problem is formulated as the following question: what is the most likely value of $Y_*$ given data $X_i$, $Y_i$, and $\Delta Y_i$.
To that end we form a joint distribution
\begin{align*}
\left[
\begin{matrix}
\bm{Y}\\ 
Y_*
\end{matrix}
\right] \sim \mathcal{N}
\left(
\begin{pmatrix}
\bm{0} \\
0
\end{pmatrix}
,
\begin{pmatrix}
K(\bm{X}, \bm{X})+ {\rm diag}(\bm{\Delta Y}) & K(\bm{X}, X_*) \\
K(X_*, \bm{X}) & K(X_*, X_*).
\end{pmatrix}^{\mathstrut}
\right) 
,
\end{align*}
where $\bm{X}$, $\bm{Y}$, and $\bm{\Delta Y}$ are the vectors composed of $X_i$, $Y_i$, and $\Delta Y_i$, which makes $K(\bm{X}, \bm{X})$ a matrix composed of $k(X_i, X_j)$.
From this, we find that $Y_*$ is normally-distributed with mean
\[
\overline{Y_*} = K(X_*, \bm{X}) [K(\bm{X},\bm{X}) + {\rm diag} (\bm{
\Delta Y})]^{-1}\bm{Y}, 
\]
and variance
\begin{equation}\label{eq:cov}
\operatorname{Var}(Y_*) = K(X_*, X_*) - K(X_*, \bm{X}) [K(\bm{X}, \bm{X}) + {\rm diag}(\bm{\Delta Y})]^{-1} K(\bm{X}, X_*).
\end{equation}

The variance of a nonlinear functional $\mathcal{F}(S)$, which is needed to predict the uncertainty of determining the melting point \eqref{eq:trans_Y}, is derived in Appendix \ref{var} in the limit of a large amount of data (in which can we can linearize $\mathcal{F}(S)$ around the mean prediction $\overline{S}$).

For a given application, one needs to find the right values of hyperparameters $\bm{\theta} = (\theta_T, \theta_V)$ in \eqref{eq:kernel-example}.
This is done by maximizing the so-called marginal likelihood $p(\bm{Y} | \bm{X}, \bm{\theta})$ which is proportional to the probability that the underlying data is distributed according to the hyperparameters $\bm{\theta}$.
The marginal likelihood is calculated according to the formula
\begin{equation}\label{eq:MG}
\begin{split}
& \log p (\bm{Y} | \bm{X}, \bm{\theta}) = -\frac 1 2 \bm{Y}^{T} [K(\bm{X}, \bm{X})+ {\rm diag}(\bm{\Delta Y})]^{-1} \bm{Y} \\
& - \frac 1 2 \log |K(\bm{X}, \bm{X}) + {\rm diag}(\bm{\Delta Y})| - \frac  n 2 \log (2 \pi),
\end{split}
\end{equation}
where $|K|$ denotes the determinant of the matrix $K$ and $n$ is the dimension of the model (number of input points).

The Bayesian variance expressed in equations \eqref{eq:cov} and \eqref{eq:V} does not directly depend on $\bm{Y}$. This allows us to define the quantity
\begin{equation}
\mathcal{H}(Q,X^*) = - \log \frac {\mathbb{V}(Q| \bm{X} \cup X^*)} {\mathbb{V}(Q| \bm{X})},
\label{eq:inf}
\end{equation}
which expresses an expected improvement of the variance of $Q$ after adding a new point $X^*$ to the dataset $\bm{X}$.
We will call \eqref{eq:inf} the information function.
By maximizing $\mathcal{H}$ with respect to $X^*$, we find the point that is best in reducing the variance of the quantity $Q$. This gives rise to our active sampling algorithm, whose essence is to simply greedily add points with maximum information to the training set, one by one.
For simplicity, when deciding which new point to add to the training set, we assume that we would add data with zero variance.


\subsection{Illustrative example: performing integration with a Gaussian process}

Before applying Gaussian processes to reconstruct the free energy function from its derivatives, let us consider an illustrative problem of simply reconstructing $f(x) = \sin(x)$ on the interval $[0,\pi]$ from its value $f(x)=0$ and nine noisy values of $f'(x)$.
To that end, we take nine evenly spaced points, $x_1=0$, $x_2=\pi/8$, \ldots, $x_9=\pi$ and consider $y_i = f'(x_i) + {\mathcal N}(0,\sigma^2)$ (i.e., added to $f'(x_i)$ some normally distributed noise) with variance $\sigma = 0.1$.
The added noise simulates the statistical noise always present in averages taken over finite MD trajectories.

\begin{figure}[h!]
	\noindent\centering{%
		\hfill
		\includegraphics{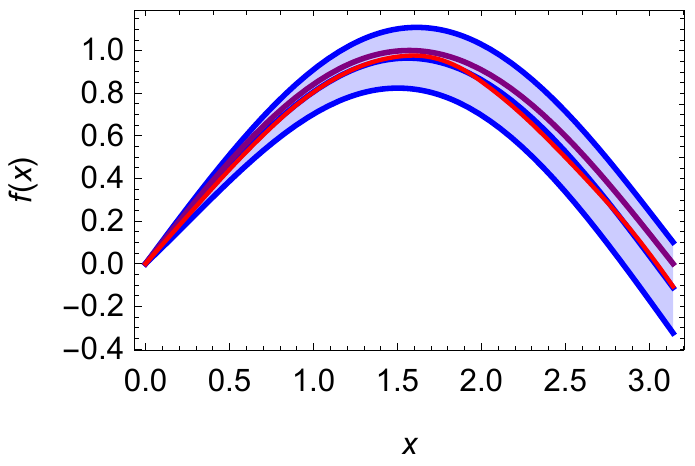}
		\hfill
		\hfill
		\includegraphics{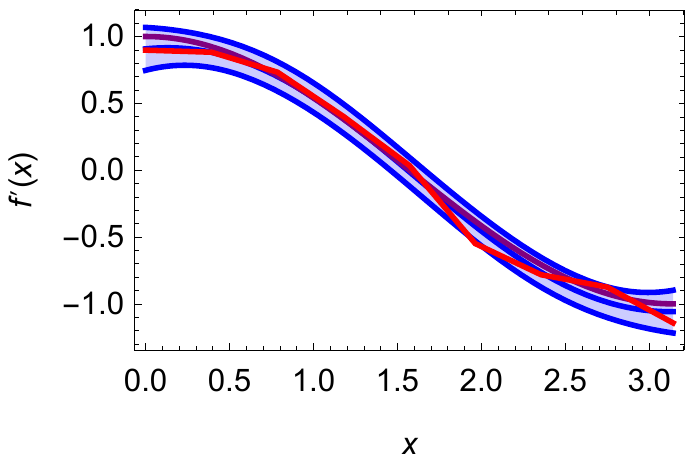}
		\hfill$\mathstrut$
	}%
	\caption{%
		Illustration of integration via the trapezoidal quadrature rule (red) and Gaussian process (blue) on noisy data.
		The left graph shows how the function $f(x) = \sin(x)$ is reconstructed and the right one shows how the derivative is reconstructed.
		The Gaussian process has the same accuracy, but produces a smooth result and also yields uncertainties of prediction within which the exact solution (purple) falls.} 
	\label{fig:1d-illustration}
\end{figure}

The results of the comparison are shown in Figure \ref{fig:1d-illustration}.
To mimic thermodynamic integration, the state-of-the-art method used for calculating the free energy, we use the second-order trapezoidal quadrature rule, shown with red in the figure.
The ``integration'' with the Gaussian process was done as outlined in Section \ref{sec:GP}.
We see that the accuracy of the two methods is comparable, but the Gaussian process yields smoother results and also gives an accurate confidence interval.

\section{Results and Discussion \label{sec:res}}

\subsection{Methods \label{sec:impl}}

The formulas from Section \ref{sec:free_en} apply to any data regardless of their source.
In this work, we use classical molecular dynamics (MD) implemented in the LAMMPS package \cite{LAMMPS} to generate the input data for the Gaussian process. Molecular dynamics simulations are performed in the canonical ensemble (NVT).

The core of classical MD simulations is the interatomic potential.
An interatomic potential is a functional form that allows one to calculate the potential energy of the system. In our work, we study the behavior of the systems described by the Lennard-Jones potential, the soft-core potential (being simply the repulsive term of the Lennard-Jones potential), and the Moment Tensor Potential \cite{shapeev2016moment,gubaev2019-alloys}.
For each of these functional forms, the total potential energy $E(\bm{x})$ can be partitioned into the sum of atomic contributions 
\[
E(\bm{x}) = \sum_i E(\bm{x}_i),
\]
where $\bm{x}_i$ is the coordinate of the atom with the index $i$. The potential energy of a particle, $E(\bm{x}_i)$, is a sum of atomic contributions within finite sphere of radius $r_{\rm cut}$ called cutoff radius
\[
E(\bm{x}_i) = \sum_{j : |\bm{x}_i - \bm{x}_j| < r_{\rm cut}} \varphi(|\bm{x}_i-\bm{x}_j|),
\]
where $\varphi$ is the function that describes potential energy of two-particle interaction with respect to their positions $\bm{x}_i$ and $\bm{x}_j$.
The potential energy of the soft-core and Lennard-Jones systems can be explicitly decomposed into a sum of pair contributions. Here and in what follows, the functional form of the potential energy of single pair interaction will be denoted by $\varphi (r)$.

\begin{figure}[h!]
	\noindent\centering{\includegraphics[width=150mm]{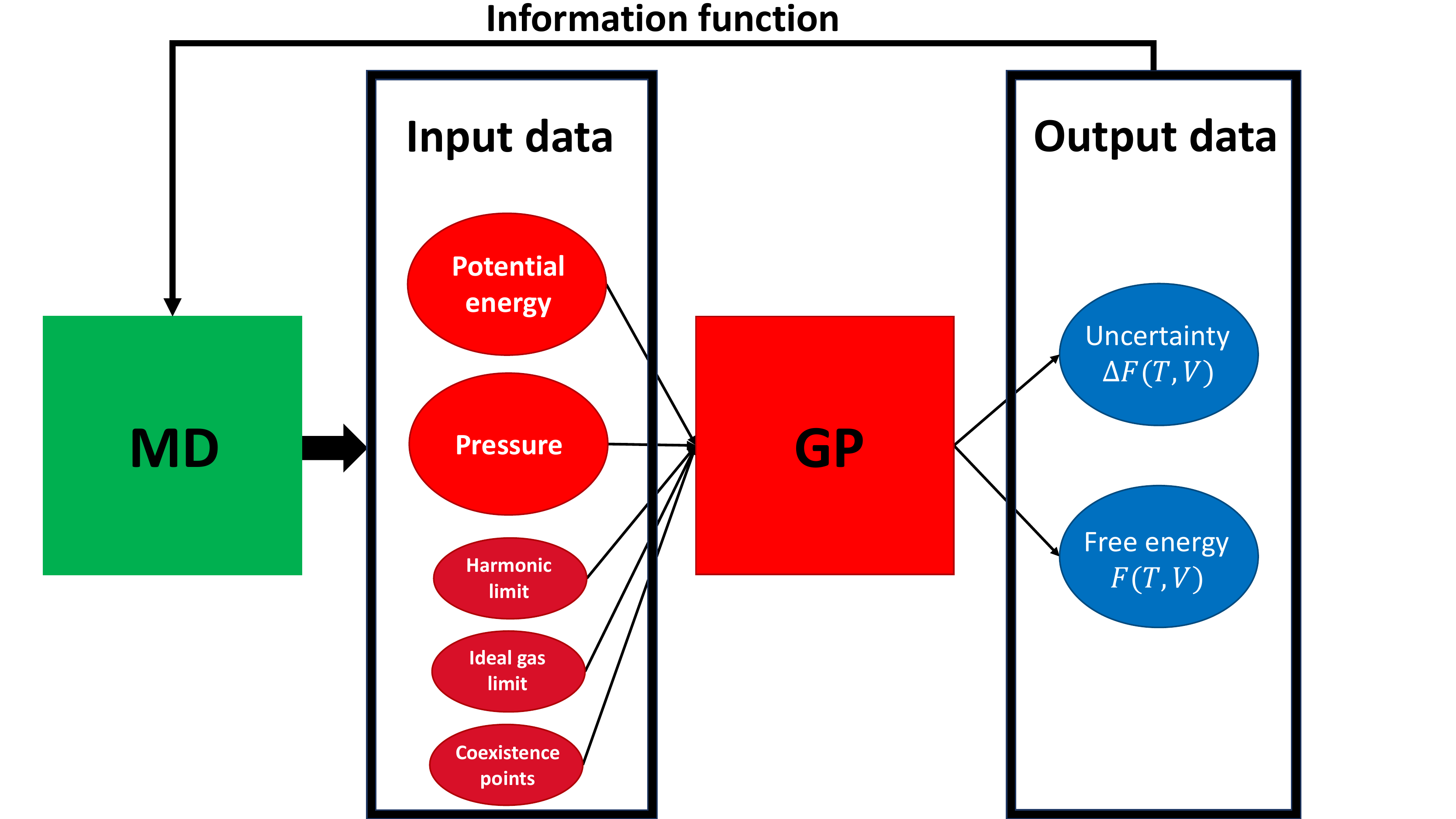}}
	\caption{General scheme of the Bayesian framework for calculating phase diagrams.} 
	\label{fig:gen_scheme}
\end{figure}

The temperature $T$, specific volume $V$, cutoff radius of interatomic potential $r_{\rm cut}$, and a number of atoms in the simulation cell $N$ form the entries of the matrix $\bm{X}$ of the Gaussian process.
The entries of the vector $\bm{Y}$ are formed from derivatives and values of the free energy. 
The derivatives are calculated via formulas \eqref{eq:e_dir_sol}, \eqref{eq:p_dir_sol}, \eqref{eq:p_dir_liq}, and \eqref{eq:e_dir_liq} using the averaged potential energy $\<E\>$ and the virial pressure $\<P\>$ obtained from MD.
An additive constant of the free energy is set via harmonic \eqref{eq:F_ref_sol} and ideal gas \eqref{eq:F_ref_liq} limits.
Where these limits are not applicable, we determine the additive constant difference via \eqref{eq:trans_Y}.
The standard deviation of the corresponding thermodynamic quantities over the MD trajectories forms entries of the vector $\bm{\Delta Y}$.  An active sampling algorithm defined in Section \ref{sec:GP} is applied to reduce the uncertainty of the target quantities systematically.
The general scheme of our framework is shown in Figure \ref{fig:gen_scheme}. 

\subsubsection*{Analysing and accelerating convergence with a Gaussian process}

Gaussian processes will allow us to automate one more job that is traditionally done manually: the analysis and acceleration of the convergence with respect to numerical parameters, such as the number of atoms or the cutoff radius.
The fundamental property of Gaussian processes enabling this is what we have already used for the derivatives: taking the limit of a Gaussian process (e.g., with $N\to\infty$), being a linear operation, still yields a Gaussian process. Moreover, we choose the kernel functional form so that it reflects as much as possible the physical behavior of the system of interest, as we illustrate in the applications below. As a result, the trained Gaussian processes reproduce free energy with high accuracy.

\subsection{Model system: soft-core potential}\label{scp_res}

We first validate our methodology on two model systems: soft-core and Lennard-Jones potentials. In the soft-core potential, there is only one phase transition.
This simplifies the comparison between our GP estimation and the existing data.
MD calculations in the case of soft-core and Lennard-Jones potentials are performed with a Metropolized Langevin thermostat \cite{besag1994comments} to avoid error related to a finite time step. The data presented below is given in the reduced Lennard-Jones units.

We consider the soft-core interatomic potential with pair interaction described by
\[
\varphi(r) = \frac{4}{r^{12}},
\]
where $r$ is the distance between two atoms.

Thanks to the simplicity of the soft-core potential, the free energy admits the following invariant transformation:
\[
F(T,V) = - T \ln V   + F (T V^4, 1).
\]
Hence, the free energy dependence for such a potential can be expressed as  $F(T, V) \sim F(T, 1)$. This means that given a dependence of the free energy on the temperature at a certain volume, we know the dependence of the free energy in arbitrary volume and temperature range.

With this invariance in mind, and also asymptotic behavior of the free energy for $V\to 0$, we  
define the GP kernel---a functional form that we use to estimate the correlation between the training set points $X_1$ and $X_2$:
\begin{align*}
k_{\rm scp}(X_1, X_2) \sim \exp\left(-\frac {\left(\left(1 - \frac {T_1} {1+T_1}\right)^{\frac 1 4} - \left(1 - \frac {T_2} {1+T_2}\right)^{\frac 1 4}\right)^{2}} {2 \theta_{T}^2} - \frac {\left(\frac {T_1} {1+T_1} - \frac {T_2} {1+T_2}\right)^2} {2 \theta_{T}'^2}\right) \\
\left(1+\frac {\theta_c^{10}} {c_1^5 c_2^5} \exp \left( -\left(\frac 1 {c_1^2}  - \frac 1 {c_2^2} \right)^2 \frac{\theta_c^4} {2} \right)\right) \exp \left( -  \left(\frac 1 {N_1}  - \frac 1 {N_2}\right)^2 \frac{\theta_N^2} {2} \right),
\label{kernel_scp}
\end{align*}
where ${\bm \theta} = (\theta_{T}, \theta_{T}', \theta_c, \theta_N)$ are the hyperparameters of our model; $T_1$, $T_2$ are the temperatures of the first and second point; $c_1$, $c_2$ and $N_1$, $N_2$ are the interatomic potential cutoff radius and the number of atoms in the simulation cell, correspondingly.
The choice of the kernel reflects our knowledge of the system in the following way.
We operate with the rescaled temperature $\frac T {1 + T}$ instead of $T$ to avoid divergence at an infinite temperature limit in the case of liquid.
We modify the temperature dependence with $\frac 1 4$ power term to account for the two-particle interaction in the liquid free energy at high temperature. The cutoff term depends on $r_{\rm cut }$ as $c^{-5}$ (in fact, the dependency in the case of the soft-core potential is stronger, but we choose the form with $c^{-5}$ to match the asymptotics for the Lennard-Jones system). Dependencies of lower order are taken into account via long term correction. Finally, the free energy dependence with respect to the number of atoms in the simulation cell is derived from Taylor series expansion and is proportional to $\frac 1 N$.

We compare our results of melting point calculation with coexistence simulation data at P = 1. The melting temperature predicted via GP is equal to $0.1849(8)$ (error in parenthesis corresponds to the 68\% confidence interval, $0.1849(8)$ should be read as $0.1849\pm0.0008$). The value obtained by our coexistence simulations is $0.1846(2)$.
The obtained data are in agreement with respect to a given confidence interval.
We note, however, that the GP provides a more reliable confidence interval being a natural statistical estimation of uncertainty, unlike the coexistence simulations in which it is not easy to control all sources of error.

\begin{figure}[h!]
\noindent\centering{\includegraphics[width=150mm]{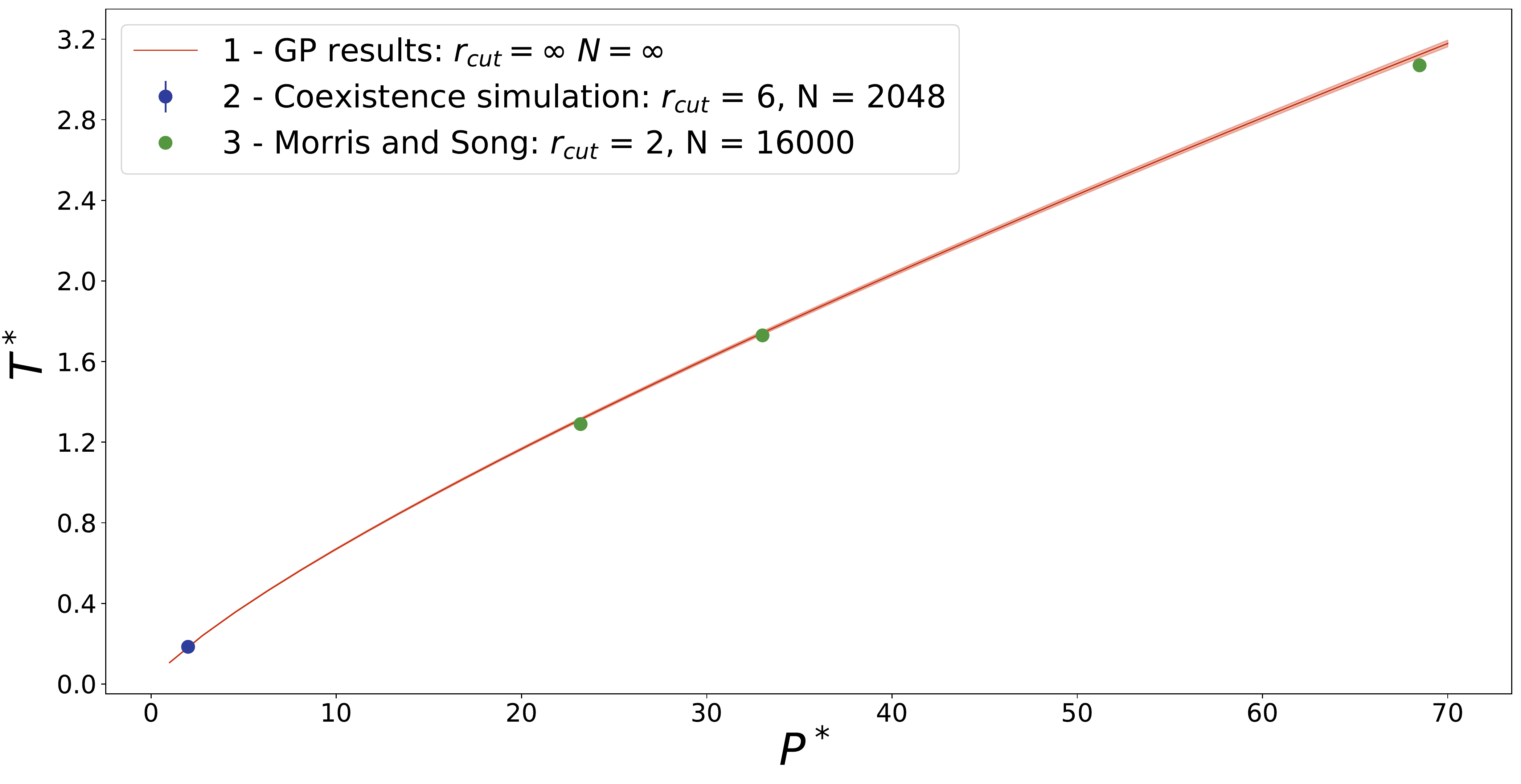}}
\caption{Melting line of the soft-core potential plotted with the 68\% confidence interval: the red curve represents the GP prediction of the melting curve at the limit of infinite cutoff and an infinite number of atoms; the blue point is our coexistence simulation result at the cutoff of 6 and number of atoms of 2048 for each phase. Green points are the results from \cite{morris2002melting}, with the cutoff of 2 and the number of atoms of 16000.
Our data is seen to be in an excellent agreement with coexistence simulation results and previous studies.}
\label{fig:fig_2}
\end{figure}

In the case of melting curve estimation, we compare our results with a direct calculation based on coexistence simulation and previous results from \cite{morris2002melting}. The obtained data is shown in Figure \ref{fig:fig_2}.
In this case, the GP estimation of the melting curve agrees well with our simulations.
Our data is in a good agreement with the previous results.
The melting point at $P\approx 70$ can be seen to deviate by about $3\sigma$ (i.e., three standard deviations), which is statistically significant, but we attribute this to the low value of cutoff ($r_{\rm cut}=2$ for the green points taken from \cite{morris2002melting}).

\subsection{Model system: Lennard-Jones potential}

The Lennard-Jones potential was selected as a system with three phases and hence a more complex phase diagram. Also, the $\frac 1 {r^6}$ term of the Lennard-Jones potential results in the strong dependence of obtained data on the size of the simulation cell and cutoff radius of the interatomic potential.
As in the soft-core case, the MD calculations are performed with a Metropolized Langevin thermostat \cite{besag1994comments}.

The potential energy of the pair interaction of the Lennard-Jones system has a form
\[
\varphi(r) =4  \left( \frac {1} {r^{12}} - \frac {1} {r^6}\right).
\]
Compared to the soft-core potential the Lennard-Jones potential has an additional term $\frac {1} {r^6}$. This leads to a phase diagram with three different phases: solid, gas, and liquid. The last two can be described by a single free energy curve because they are indistinguishable at temperatures above critical. The kernel for the Lennard-Jones free energy is similar to the kernel of the soft-core potential except that we treat the volume dependence of the free energy explicitly in the kernel
\begin{align*}
k_{\rm LJ}(X_1, X_2) \sim k_{\rm scp}(X_1, X_2) \left( \exp \left(- \frac {\left( \rho_1 - \rho_2 \right)^2} {2 \theta_{\rho}^2}\right)  - \exp \left( -\frac {\left( \rho_1^2 + \rho_2^2 \right)} {2 \theta_{\rho}^2}\right) \right),
\end{align*}
where $\rho_i$, simply defined as $\rho_i := V_i^{-1}$, are densities at first and second points.
We subtract $\exp \left( -\frac {\left( \rho_1^2 + \rho_2^2 \right)} {2 \theta_{\rho}^2}\right)$ from the Gaussian kernel to account for the zero-density (or infinite-volume) limit of the free energy \eqref{eq:liq_lim}.

We first validate our algorithm by computing the critical and triple points of the Lennard-Jones potential.
The critical point is the point on the phase diagram where the liquid and gas phases become indistinguishable. Our results, along with the previous studies, are shown in Table \ref{tab:cp}. 
\begin{table}[h!]
\begin{center}
\caption{Critical point (index ${\rm crit}$) estimation. $r_{\rm cut}^*$ denotes the cutoff radius of the interatomic potential with a long-range correction.
	Our results are in a good agreement with the previous studies.}
\begin{ruledtabular}
\begin{tabular}{@{}ccccc}
Source           & $r_{\rm cut}^*$ & System size  & $T_{\rm crit}$ & $\rho_{\rm crit}$ \\ \hline
ref.\ \cite{caillol1998critical} & $\infty$    & $\infty$ & 1.326(1)   & 0.316(1)      \\
This work  & $\infty$   & $\infty$ & 1.327(1)   & 0.316(2)  \\ 
\end{tabular}
\end{ruledtabular}
\label{tab:cp}
\end{center}
\end{table}
There are a lot of studies for comparison, but we have chosen \cite{caillol1998critical} as the one with the most reliable values. In \cite{caillol1998critical}, the authors calculated the critical point using MC simulation and estimated effect of the system's finite size. Our results agree well with respect to confidence interval. Moreover, the estimated confidence interval is in agreement with the previous results.

\begin{figure}[h!]
\noindent\centering{\includegraphics[width=145mm]{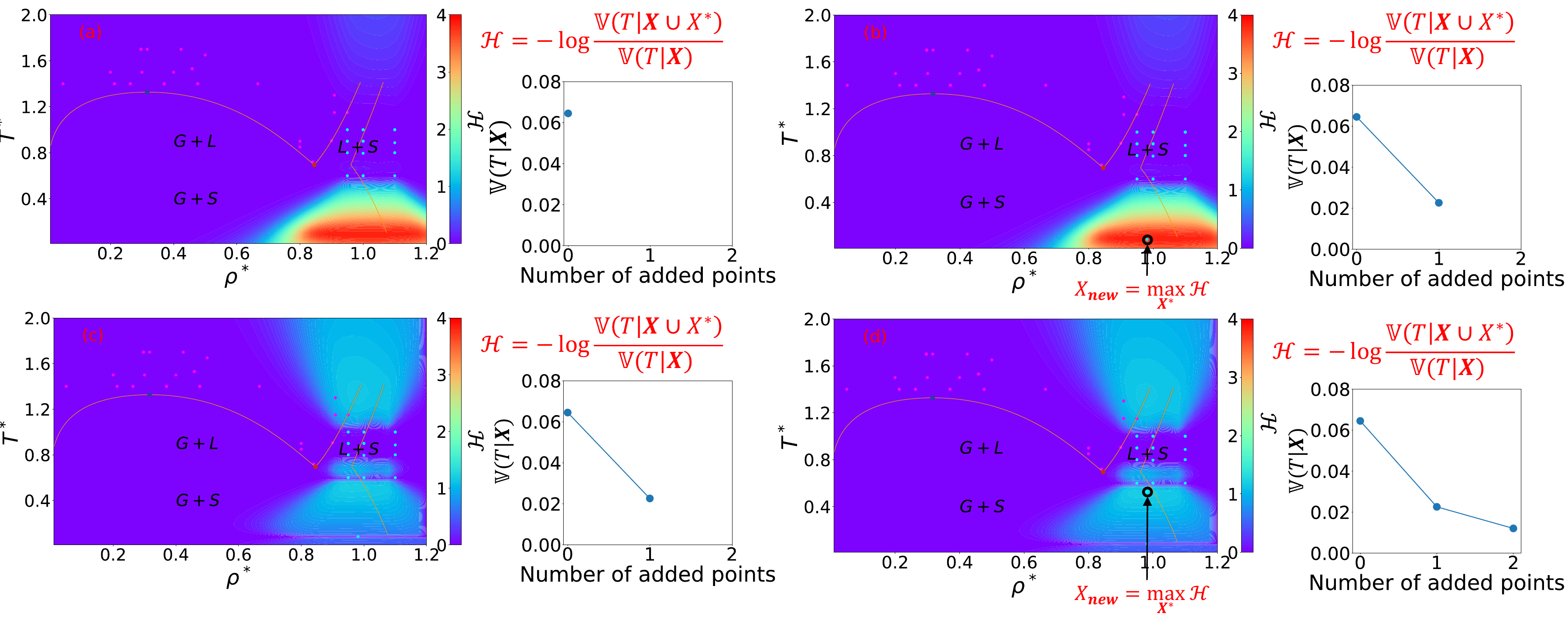}}
\caption{Illustration of the active sampling algorithm applied to the calculation of the triple point temperature estimation. Color plot represents the value of the information function $\mathcal{H}$ from \eqref{eq:inf} at a given point of the phase diagram.
	In step (a), the information function is calculated at each point.
	Then, in step (b), a new point $X_{\rm new}$ that maximizes the information function is added to the dataset.
	This finalizes the first iteration of the algorithm.
	Figures (c) and (d) correspond to the second iteration of the active sampling procedure. The proposed active sampling algorithm allows one to decrease the error of the target property systematically.}
\label{fig:AL}
\end{figure}

We next estimate the triple point as a part of our validation procedure.
An active sampling algorithm was used to improve the accuracy of the triple point calculation significantly.
An illustrative example of two steps of the active sampling algorithm is shown in Figure \ref{fig:AL}.

Our results of the triple point calculation, along with previous studies, are presented in Table \ref{tab:tp_res}. Our results are in perfect agreement with the work \cite{mastny2007melting}, where the authors performed the study of convergence with respect to the size of the system. From the comparison of the results with \cite{ahmed2009solid}, one can deduce that the choice of the $r_{\rm cut}$ strongly affects the triple point calculation for the Lennard-Jones potential. 
\begin{table}[h!]
\begin{center}
\caption{Triple point (index ${\rm tp}$) estimation. Indexes ``gas'', ``liq'', and ``sol'' denote gas, liquid, and solid phases respectively. Our results agree well with the existing studies.}
\begin{ruledtabular}
\begin{tabular}{@{}ccccccc}
Source          &$r_{\rm cut}^*$  & System  & $T_{\rm tp}$   & $\rho_{\rm gas} $ & $\rho_{\rm liq}$ & $\rho_{\rm sol}$ \\ 
                      & & size                   &                   &  $\cdot 10^{-3}$ &                     & \\ \hline
Ladd and Woodcock \cite{ladd1978interfacial}   & 2.5   & 1500      & 0.67(1)  & ...          & 0.818(4)     & 0.963(6)     \\ 
Hansen \cite{hansen1970phase} & ...   & 864        & 0.68(2)  & ...          & 0.85(1)      & ...          \\ 
Kofke \cite{kofke1993direct} & ... & 236      & 0.698    & ...          & 0.854        & 0.963        \\ 
Kofke \cite{kofke1993direct}  & ... & 932      & 0.687(4) & ...          & 0.850        & 0.960        \\ 
Ahmed and Sadus \cite{ahmed2009solid}   & 2.5  & 2048     & 0.661    & ...          & 0.864        & 0.978        \\ 
Mastny and Pablo \cite{mastny2007melting} & 6   & $\infty$ & 0.694(4) & ...          & ...          & ...          \\ 
This work  & $\infty$   & $\infty$ & 0.695(4) &  1.9(1)          &    0.845(2)       & 0.961(1)          \\ 
\end{tabular}
\end{ruledtabular}
\label{tab:tp_res}
\end{center}
\end{table}

\begin{figure}[h!]
\noindent\centering{\includegraphics[width=150mm]{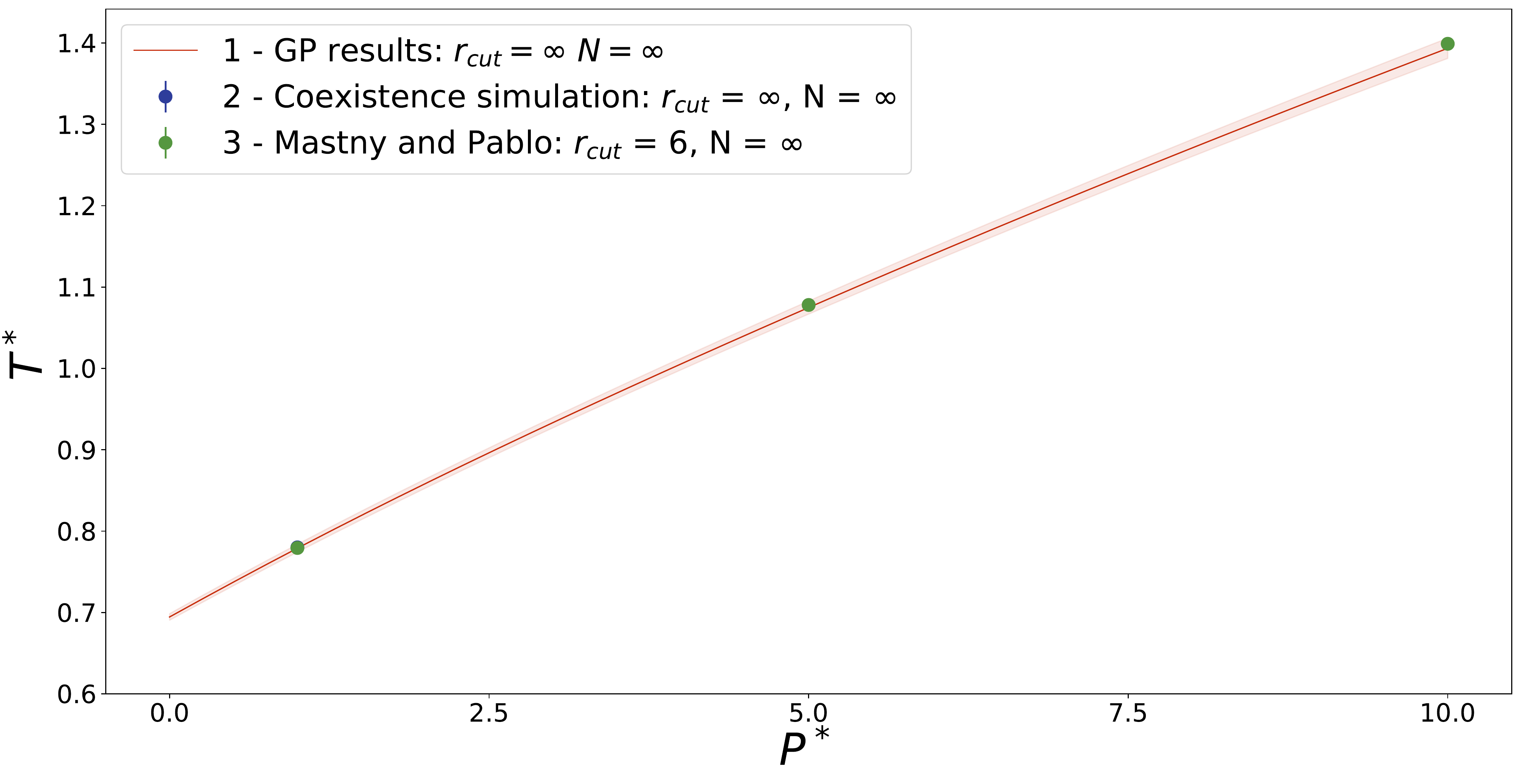}}
\caption{Melting curve of the Lennard-Jones potential. The red curve corresponds to the calculation via GP. Rhombus is an extrapolation of coexistence simulation results to the limit of infinite $r_{\rm cut}$ and infinite system size. Green dots denote the existing results from \cite{mastny2007melting}. Our results are in an excellent agreement with \cite{mastny2007melting} and coexistence simulation data.} 
\label{fig:fig_4}
\end{figure}

We next compare the melting line of the Lennard-Jones potential with our coexistence simulation results and the previous studies.
The obtained data is shown in Figure \ref{fig:fig_4}.
Our results agree well with the previous studies. We have chosen \cite{mastny2007melting} as reference data because the authors have examined the effect of the finite system size on the triple and melting points calculation. Compared to the study \cite{mastny2007melting} we have treated cutoff and system size as explicit parameters of our model.

\begin{figure}[h!]
\noindent\centering{\includegraphics[width=150mm]{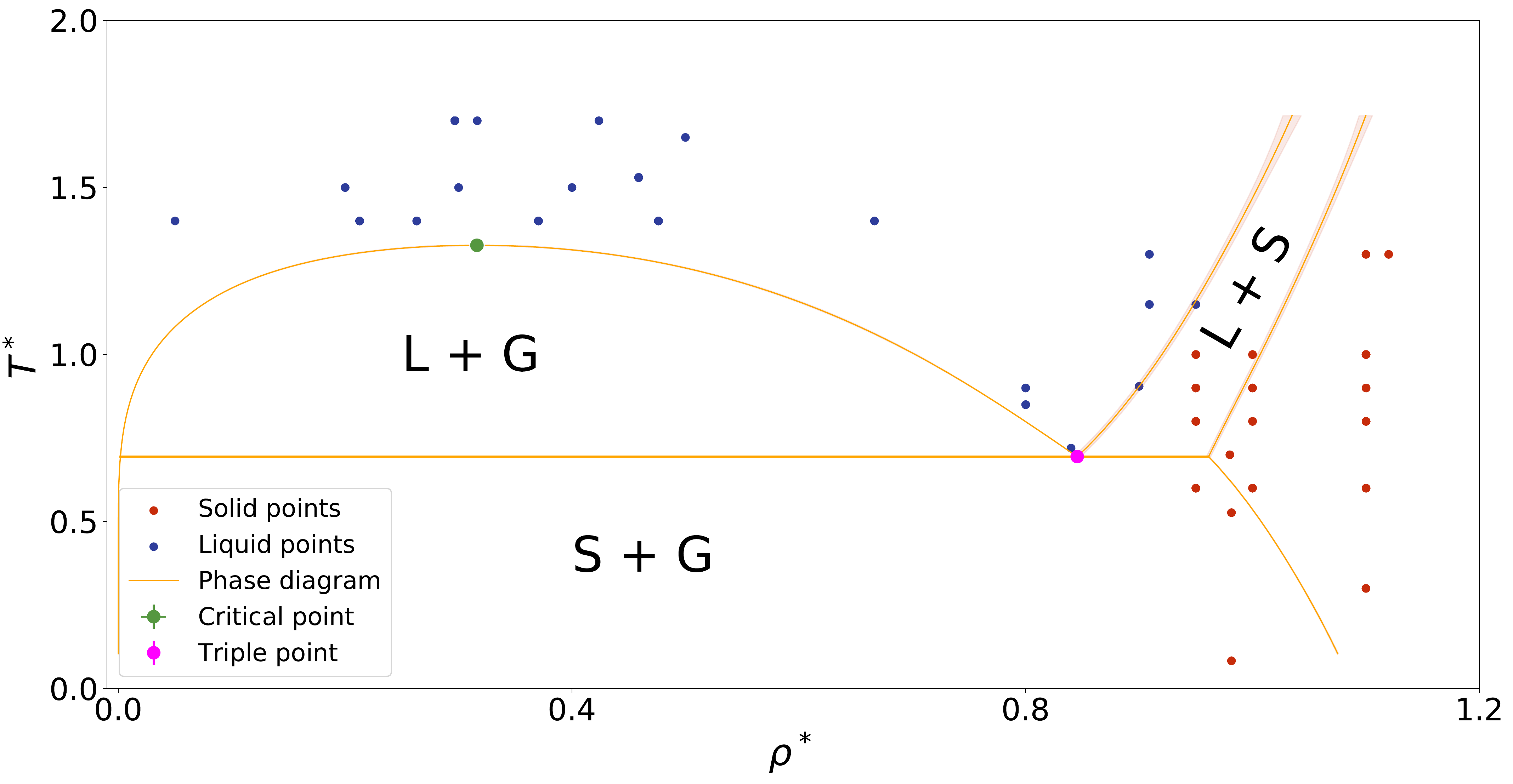}}
\caption{The Lennard-Jones phase diagram at the limit of infinite $r_{\rm cut}$ and $N$ with a confidence interval.
	Blue and red points are the training points for solid and liquid, respectively; the orange curve is a phase diagram estimated via GP.
	The green dot denotes the critical point.
	The pink dot represents the triple point.
	Only a few points of both solid and liquid phases are used to calculate all possible phase transitions of the Lennard-Jones system with high accuracy.} 
\label{fig:fig_5}
\end{figure}

We next calculate the phase diagram of Lennard-Jones at the limit of infinite cutoff radius and system size. The phase diagram is presented in Figure \ref{fig:fig_5}.
As one can see, we have predicted the phase diagram and estimated the confidence interval of our calculation. Also, the phase diagram obtained via GP fitted with a limited number of points for both phases is in agreement with the previous studies.

\subsection{Physical system: lithium}

Finally, we apply our methodology to lithium, modeled with a machine-learning potential trained on DFT calculations.
We compute the phase diagram of lithium at pressures below 30 GPa and temperatures above room temperature in this work.
We hence examine the bcc, fcc, and liquid phases of lithium.
In particular, the bcc phase is not stable at $T=0$, therefore to accurately obtain the additive constant of the bcc-Li free energy, we rely on the harmonic limit for the fcc phase, fcc-liquid coexistence simulations to obtain the additive constant of the liquid phase, and bcc-liquid coexistence simulations to finally obtain the additive constant of the bcc phase.

Thermodynamic data is obtained with the use of Moment Tensor Potential (MTP) \cite{shapeev2016moment,gubaev2019-alloys} as implemented in the MLIP software package \cite{novikov2020-mlip}.
We have trained a single MTP for all phases fitted on quantum mechanical data in the examined temperature and pressure range.
The MTP potential was actively trained on-the-fly by running MD simulations of fcc Li with 108 atoms and bcc Li with 128 atoms for a range of volumes covering the pressure range of interest and increasing temperature from normal conditions up to 900 K to observe melting.
The default, level-16 MTP potential was used, with the cutoff radius of $5$\AA\ and the {\tt mindist} value set to $1.4$ \AA.
The DFT calculations were conducted with the plane-wave DFT, and a PAW pseudopotential with one electron treated as the valence electron, as implemented in the VASP package \cite{VASP1,VASP3,VASP4}.

The kernel for the lithium fcc phase is chosen as
\begin{align*}
k_{\rm Li}^{\rm fcc}(x_1, x_2) \sim \exp\left(- \frac {(T_1 - T_2)^2} {2 \theta_T^2}\right)  \exp \left(- \frac {\left( \rho_1 - \rho_2 \right)^2}  {2 \theta_{\rho}^2 } \right) \exp \left( -  \left(\frac 1 {N_1}  - \frac 1 {N_2}\right)^2 \theta_N^2 \right).
\label{kernel_Li}
\end{align*}

For liquid and bcc phases we do not have reference at zero temperature. In order to account for divergence of entropy at this limit we modify temperature part of the kernel by adding the $\frac 1 T$ term:
\begin{align*}
k_{\rm Li}^{\rm bcc, liquid}(x_1, x_2) \sim \left(\frac {\theta_T^2} {T_1 T_2} + \exp\left(- \frac {(T_1 - T_2)^2} {2 \theta_T'^2}\right) \right) \\
\exp \left(- \frac {\left( \rho_1 - \rho_2 \right)^2}  {2 \theta_{\rho}^2 } \right) 
\exp \left( -  \left(\frac 1 {N_1}  - \frac 1 {N_2}\right)^2 \frac{\theta_N^2} {2} \right).
\end{align*}

In the cases where we do not have reference data, we explicitly add melting points to the dataset. Using the MTP fitted on quantum mechanical data, we cannot approach the limit of infinite temperature or volume (since the potential was not fitted at those conditions).
Also, the bcc phase of lithium is dynamically unstable at zero temperature. Thus we use the bcc-liquid and fcc-liquid phase transition as reference data. For this reason, to validate our approach, we compare the prediction of GP with coexistence simulation results at various phase transition points. Results are presented in Table \ref{tab:li_melt}. Predicted results lie well within the 95 \% confidence interval with respect to the coexistence simulation data.
\begin{table}[h!]
\begin{center}
\caption{Comparison between the GP-predicted phase transition temperatures and the coexistence simulation results.
	The indices ``coex'' and ``GP'' denote coexistence simulation and estimations of the GP, respectively. ``*'' refers to the phase transition points used as free energy references.
	The points not marked with ``*'' are the validation data, not used in the fitting of the GP.
	$\sigma$ refers to the total confidence interval (combined coexistence and GP confidence intervals).
	$\Delta T$ is the absolute difference (error) between melting temperature predicted by coexistence simulation and GP ($\Delta T = |T_{\rm coex} - T_{\rm GP}|)$.
	$\Delta T/\sigma$ is the error measured in standard deviations, value $1$ is the expected value of the error, a value above 2 would indicate a statistically significant error.
	Our GP-based algorithm is thus in an excellent agreement with coexistence simulation results.
}
\begin{ruledtabular}
\begin{tabular}{@{}ccccc}
Phase         &P,   & $T_{\rm coex}$,  & $T_{\rm GP}$,    & $\Delta T/ \sigma$  \\ 
trainsition   & GPa &        K   &  K &\\ \hline
fcc - liq$^*$   & 12 & 495(2)   & 495(2)      & 0 \\ 
fcc - liq   & 15 & 482(3)   & 483(2)     & 0.5  \\ 
fcc - liq   & 20 & 456(2)   & 454(2)      & 0.7 \\ 
fcc - liq   & 30 & 376.3(5)   & 378(2)      & 0.8  \\ 
bcc - liq$^*$   & 0 & 476(2)   & 476(2)      & 0 \\ 
bcc - liq   & 4 & 521(3)   & 521(2)      & 0.03 \\ 
\end{tabular}
\end{ruledtabular}
\label{tab:li_melt}
\end{center}
\end{table}
Finally, we compute the phase diagram of lithium in a range of pressures below 30 GPa, and a range of temperatures above 300 K extrapolated to the limit of infinite system size.
	To that end, we first train the MTP for these conditions by running short NVT-MD trajectories covering the chosen pressure and temperature conditions, and actively learning the quantum-mechanical interaction on-the-fly \cite{podryabinkin2017-AL,gubaev2019-alloys,novikov2020-mlip}.
	In total, 290 108-atom and 128-atom (for fcc and bcc respectively) crystalline and liquid configurations were selected and computed on DFT.
	We then switch off active learning and used the final potential with the developed methodology to construct the phase diagram of Li.

\begin{figure}[h!]
\noindent\centering{\includegraphics[width=150mm]{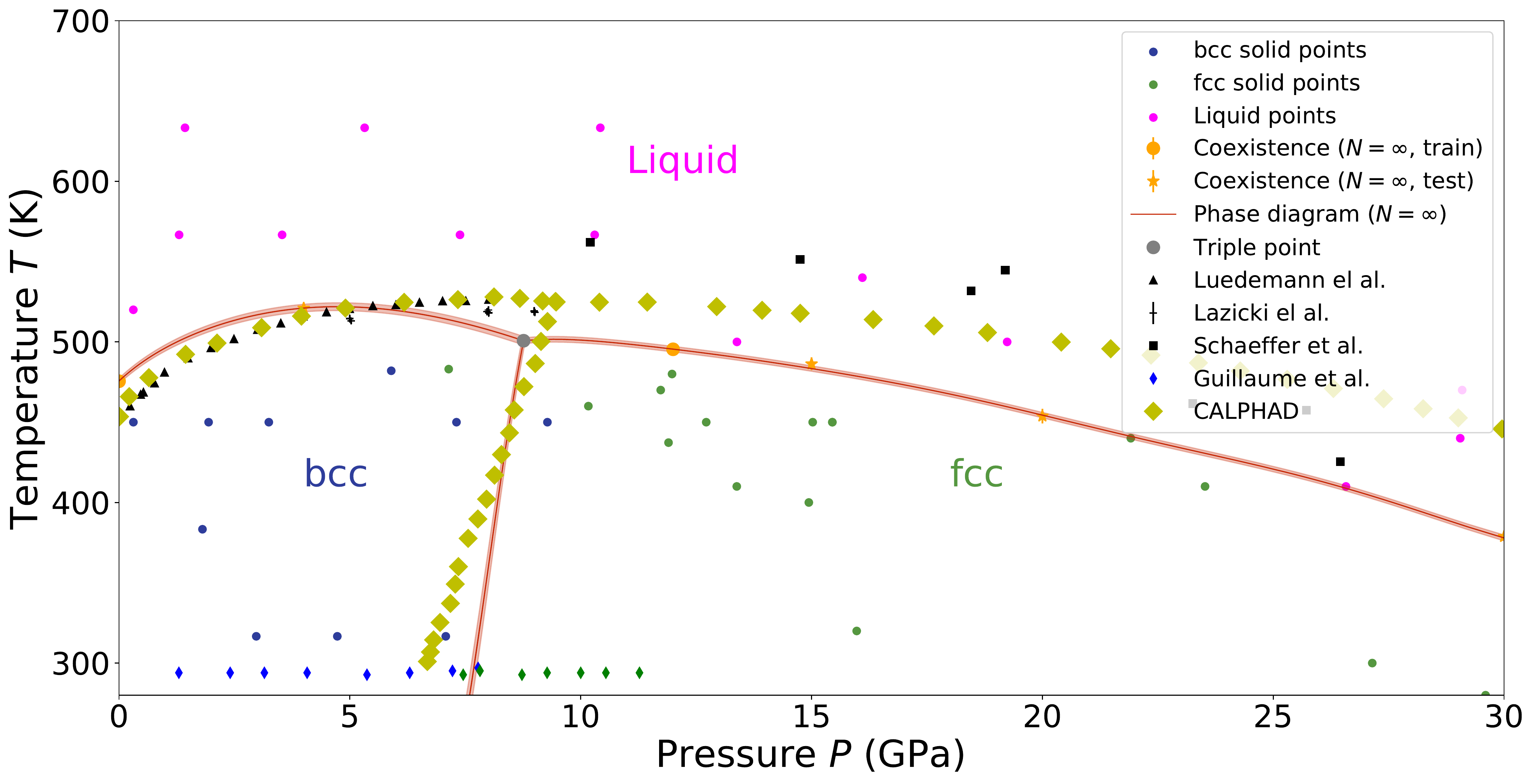}}
\caption{Lithium phase diagram at the infinite system size limit plotted with a confidence interval. Blue, green, and pink dots are training points for bcc, fcc, and liquid phases, respectively.
	The orange points are the coexistence simulation results.
	The red curve is the phase diagram obtained via GP.
	The grey dot is the location of the bcc-fcc-liquid triple point. The blue and green diamonds represent the fcc and bcc data from \cite{guillaume2011cold}.
	The experimental results \cite{luedemann1968melting,lazicki2010high,schaeffer2012high} are denoted by black markers.
	The CALPHAD approximation, based on the data from \cite{luedemann1968melting,lazicki2010high,schaeffer2012high}, is shown by diamonds.
	Overall, we observe a very good agreement with the experimental data.
} 
\label{fig:fig_7}
\end{figure}

The corresponding phase diagram, coexistence simulation results, and experimental data are shown in Figure \ref{fig:fig_7}.

We observe that the GP estimates the phase diagram with high accuracy in a wide temperature and pressure range.
Moreover, GP predicts the bcc-liquid phase transition accurately when compared to the experimental results \cite{luedemann1968melting,lazicki2010high}.
The difference between our calculation and previous studies is less than 10 K. However, the fcc-liquid phase transition shows a greater divergence from the experimental data \cite{schaeffer2012high}. The difference could be due to inaccuracies of the DFT calculations as compared to the experimental data. The fcc-bcc phase transition below 500 K and above 300 K is not explored to the authors' knowledge.
The available data around 300 K from \cite{guillaume2011cold} is in an excellent agreement with the transition estimated via GP: our bcc-fcc transition line goes directly through the region of pressures where both fcc and bcc phases were realized experimentally.

\begin{figure}[h!]
\noindent\centering{\includegraphics[width=145mm]{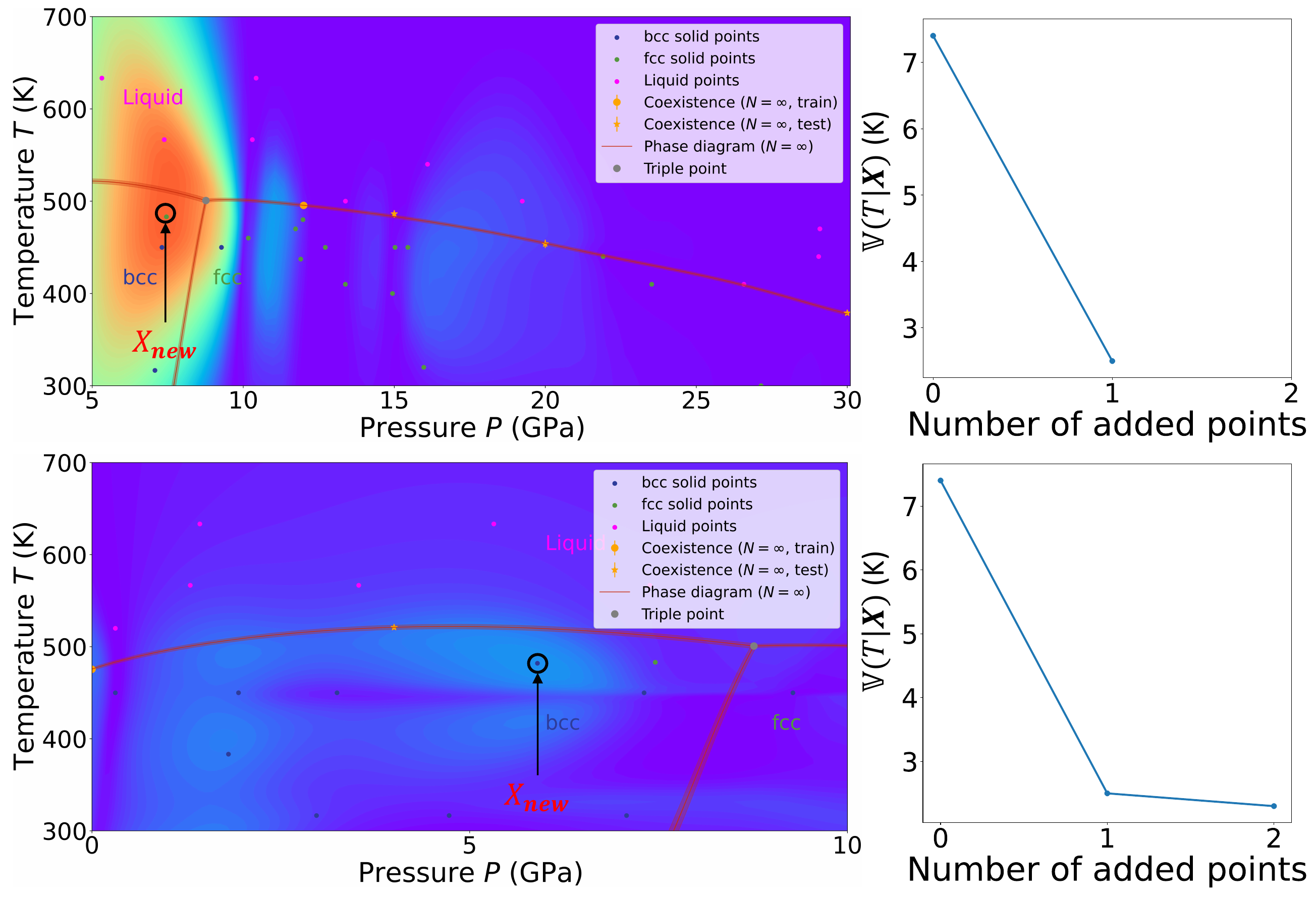}}
\caption{The active sampling algorithm applied to the calculation of the triple point temperature estimation of Lithium phase diagram. Color plot represents the value of the information function $\mathcal{H}$ from \eqref{eq:inf} at a given point of the phase diagram. The proposed active sampling algorithm allows one to optimize datasets corresponding to different phases.}
\label{fig:AL_Li}
\end{figure}

Also, we have calculated the pressure $P_{\rm tp}$ and temperature $T_{\rm tp}$ at the triple bcc-liquid-fcc point. $P_{\rm tp}$ is equal to  8.8 GPa; $T_{\rm tp} = 501(2) $ K. The obtained result is in agreement with experimental estimations. As in the case of the Lennard-Jones potential, we have applied an active learning strategy to improve the accuracy of the triple points. An example of two steps of the algorithm applied to the lithium system is given in Figure \ref{fig:AL_Li}. 

	The total computational cost of constructing the phase diagram consists of three major parts: obtaining the quantum-mechanical data for training the machine-learning potential, calculation of melting points, and running molecular dynamics to obtain the free energy derivatives from statistical averages.
	The cost of the quantum-mechanical calculations was 28\,000 CPU hours, about 100 hours per configuration (a typical calculation takes about 3 hours when parallelized over 36 cores).
	The cost of the calculation of the two melting points used in the training set of the Gaussian process was about 10\,000 hours.
	Finally, the NVT-MD calculations took 16\,000 CPU hours.
	The computational cost of the last part is comparable to the efficiency of a few melting point calculations, however, our approach allows one to fit the entire free energy surface with an error of less than 0.1 meV.

\section{Concluding remarks \label{conc}}

The construction of a phase diagram from atomistic simulation data is typically associated with a significant amount of manual work consisting of manual selection of numerical parameters, convergence tests, determining conditions at which to run simulations, analysis of the results, going back to the earlier stages if needed, etc.
In the present work, we have developed a Gaussian process-based methodology automating these stages of phase diagram calculation.
In particular, the Gaussian process allows us to reconstruct the free energy function based on various data sources (harmonic limit, zero-density limit, MD averages, and coexistence simulations).
Furthermore, the dependence of the free energy on the numerical parameters can also be learned, which allows us to converge the results and estimate the exact value automatically together with the confidence interval that includes the error of extrapolation of the results with respect to the numerical parameters. 
On top of that, the Gaussian predictive variance allows for automatic sampling algorithms, automating the job of selecting the parameters for simulations.
In view of these features of our approach, we believe that it will become increasingly useful with the rise of complex automatic protocols of calculating materials properties \cite{janssen2019pyiron}.

We have validated our algorithm on two model systems, soft-core potentials and the Lennard-Jones potentials, chosen as systems with plenty of available data in the literature.
Our results agree very well with those from the literature, deviating only in those cases when the reference data we compare with is calculated with low values of convergence parameters (such as the Lennard-Jones cutoff radius).
We then applied our methodology to lithium as an interesting example of a physical system.
Comparison to our coexistence simulations as well as the existing experimental data shows a very good agreement, with discrepancies attributed to the error of DFT itself.

\begin{acknowledgments}

A.S.\ thanks Richard Otis (Caltech) for extensive discussions that lead to the creation of this work.
This work was supported by the Russian Foundation for Basic Research under Grant.\ No.\ 20-53-12012.

\end{acknowledgments}

\bibliography{manuscript}

\appendix

\section{Derivation of Free Energy Relations}

\subsection{Reference for Solid}\label{sec:solid-ref}
Let us expand the free energy \eqref{eq:free_energy} around $T=0$.
To that end we split $\bx$ into two families of degrees of freedom: $x_N$ and $\tilde{\bx} = (x_1,\ldots,x_{N-1})$.
Let $\beta := T^{-1}$.
The determinant of the Jacobian of this transformation is $1$ and hence we can write
\begin{align*}
\beta \hat{F}_{\rm ref}(T) - \hat{S}(T) =& - \log \int {\rm d} x_N \int \exp(-\beta \hat{E}(\tilde{\bx},x_N)) {\rm d}\tilde{\bx}.
\end{align*}
Because of translational symmetry, we can fix $x_N=0$ in the inner (second) integral and hence
\begin{align*}
\beta \hat{F}_{\rm ref}(T) - \hat{S}(T) =& - \log \hat{V} - \log \int_{\hat{V}^{N-1}} \exp(-\beta E(\tilde{\bx},0)) {\rm d}\tilde{\bx}.
\end{align*}
Now note that for small $T$, only the energy near the ground state $\bx_{0}$ is relevant.
Given a ground state $\bx_0$, it is repeated $(N-1)!$ times in the integral because of permutation invariance in $E(\tilde{\bx},0)$.
Hence
\begin{align*}
\beta \hat{F}_{\rm ref}(T) - \hat{S}(T) =& - \log \hat{V} - \log (N-1)! -\log \int_{\tilde{\bx} \sim \bx_0} \exp(-\beta E(\tilde{\bx},0)) {\rm d}\tilde{\bx},
\end{align*}
where the integration is taken specifically around the ground state $\bx_0$.
$E(\tilde{\bx},0)$ has a nondegenerate Hessian around the ground state around which we will expand
\begin{align*}
\beta \hat{F}_{\rm ref}(T) - \hat{S}(T) =& - \log \hat{V} - \log (N-1)! + \beta \hat{E}_0
\\&-\log  \int \exp(-\<\bx | \tilde{H} | \bx\>/(2 T)) {\rm d}\bx + O(T),
\end{align*}
where $\tilde{H}$ is the Hessian of $E(\tilde{\bx},0)$ at the ground state.
Carrying out the integration and calculating yields
\begin{align*}
\beta (\hat{F}_{\rm ref}(T) -\hat{E}_0) - \hat{S}(T)
=&
- \log \hat{V} - \log (N-1)!
\\&-
\log \big((2\pi T)^{3N - 3} / \det\tilde{H}\big)^{1/2}
+ O(T)
\\=&
- \log \hat{V} + \log(N) - \log N!
\\&
-{\textstyle\frac{1}{2}} (3N-3) \log (2\pi T)
+{\textstyle\frac{1}{2}} \log \det\tilde{H}
+ O(T)
\\=&
- \log \hat{V} + \log(N) - N \log N + N - \log(2\pi N) + O(N^{-1})
\\&
-{\textstyle\frac{1}{2}} (3N-3) \log (2\pi T)
+{\textstyle\frac{1}{2}} \log \det\tilde{H}
+ O(T)
\\=&
N \big(
-\log N + 1 -{\textstyle\frac{3}{2}} \log (2\pi T)
+{\textstyle\frac{1}{2 N}} \log \det\tilde{H}
\big)
\\&
\big(- \log \hat{V} + {\textstyle\frac{1}{2}} \log(2\pi)
+ {\textstyle\frac{3}{2}} \log (T) \big)
+ O(T + N^{-1}).
\end{align*}
We can now make the reverse change of variables from $\tilde{\bx}$ back to $\bx$, with $\det\tilde{H} = \det\hat{H}$, where $\hat{H}$ is the $(3N-3)\times(3N-3)$ Hessian of $\hat{E}$ computed at the ground state and projected onto the subspace orthogonal to the center of mass $x_{\rm c} = \frac{1}{N} \sum_i x_i$.
And finally, we will use the simplified version of this formula for the intensive quantities
\begin{align*}
\beta (F_{\rm ref}(T)-E_0) - S(T)
=
-\log N + 1 -{\textstyle\frac{3}{2}} \log (2\pi T)
+{\textstyle\frac{1}{2 N}} \log \det\hat{H}
+ O(T + N^{-1}).
\end{align*}

Hence for solid, we choose
\[
F_{\rm ref}(T) := E_0 + T \big( -\log (NV) + 1 -{\textstyle\frac{3}{2}} \log (2\pi T) \big),
\]
which is the same as \eqref{eq:F_ref_sol}, and then \eqref{eq:entropy_sol} follows.
We note that we could leave out $V$ from $F_{\rm ref}$, but then it would enter the expression for ${{\rm \partial}S}/{{\rm \partial}V}$ and hence create asymmetry with the liquid.

\subsection{Derivative for Solid}\label{sec:solid-der}

An NVT-thermostatted molecular dynamics used in this work produces the averages
\[
\< f \> := \frac{\int f(\bx) \exp(-\beta \hat{E}(\bx)) {\rm d}\bx}{\int \exp(-\beta \hat{E}(\bx)) {\rm d}\bx}.
\]
Here the integral is over $\hat{V}^N$, and we omit the region of integration when it is clear from the context.

Let us use this formula to find
\begin{align*}
	\frac{{\rm \partial}(\beta F_{\rm ref} - S)}{{\rm \partial}\beta}
	&=
	-\frac{{\rm \partial}}{{\rm \partial}\beta} \log \int \exp(-\beta E(\bx)) {\rm d}\bx
	\\ &=
	-\Big(\int \exp(-\beta E(\bx)) {\rm d}\bx\Big)^{-1}
	\int (-E(\bx)) \exp(-\beta E(\bx)) {\rm d}\bx.
\end{align*}

Then we can determine the relation between the mean potential energy and the free energy: 
\begin{align*}
	\<E\>
	&= \frac{{\rm \partial}(\beta F_{\rm ref} - S)}{{\rm \partial}\beta}
	= -T^2 \frac{{\rm \partial}(\beta F_{\rm ref}-S)}{{\rm \partial}T}.
\end{align*}

Hence
\begin{align*}
	-\frac{{\rm \partial}S}{{\rm \partial}\beta}
	&=
	-\frac{{\rm \partial}(\beta F_{\rm ref})}{{\rm \partial}\beta} + \<E\>
	\\&=
	-\frac{{\rm \partial}}{{\rm \partial}\beta} \big(
	\beta E_0 -\log N + 1 -{\textstyle\frac{3}{2}} \log (2\pi T)
	\big) + \<E\>
	\\&=
	-\frac{{\rm \partial}}{{\rm \partial}\beta} \big(
	{\textstyle\frac{3}{2}} \log (\beta)
	\big) + \<E-E_0\>
	=
	\<E-E_0\>- {\textstyle\frac{3}{2}} T,
\end{align*}
or
\begin{align}
	\frac{{\rm \partial}S(T, V)}{{\rm \partial}T} 
	&=
	T^{-2} \<E-E_0\> - {\textstyle\frac{3}{2}} T^{-1}.
	\label{e_dir_sol_temp}
\end{align}

Let us find a similar expression for the partial derivative with respect to volume $V$.
\begin{align*}
	\frac{{\rm \partial}(\beta F_{\rm ref}-S)}{{\rm \partial}V}
	&=
	-\frac{{\rm \partial}}{{\rm \partial}V} \log \int \exp(-\beta E(\bx)) {\rm d}\bx
	\\ &=
	-\Big(\int \exp(-\beta E(\bx)) {\rm d}\bx\Big)^{-1}
\int \beta\left(-\frac{{\rm \partial} E(\bx)} {{\rm \partial}V}\right)\exp(-\beta E(\bx)) {\rm d}\bx
	-1/V
	.
\end{align*}

Assuming that $-\frac{{\rm \partial} E(\bx)} {{\rm \partial}V} = P(\bx)$, the formula can be rewritten as
\begin{align*}
	\frac{{\rm \partial}(\beta F_{\rm ref}-S)}{{\rm \partial}V}
	&=
	-\Big(\int \exp(-\beta E(\bx)) {\rm d}\bx\Big)^{-1}
	\int \beta P(\bx) \exp(-\beta E(\bx)) {\rm d}\bx
	-1/V
	.
\end{align*}

From this, we can derive the relation between the mean full pressure (sum of the ideal and virial parts) and free energy
\begin{align*}
	\<P\> = -\frac 1 {\beta} \frac{{\rm \partial}(\beta F_{\rm ref}-S)}{{\rm \partial}V}
	-\frac{1}{\beta V}.
\end{align*}

Hence
\begin{align*}
	-\frac{{\rm \partial}S}{{\rm \partial}V} &=  -\frac{{\rm \partial}(\beta F_{\rm ref} )}{{\rm \partial}V} - \beta \<P\>
	\\&=
	-\frac{{\rm \partial}}{{\rm \partial}V} \big(
	\beta E_0 -\log (N(V)) + 1 -{\textstyle\frac{3}{2}} \log (2\pi T)
	\big) - \beta \<P\>
	-\frac{1}{V}
	\\& = \beta ( P_0 - \<P\> ),
\end{align*}
or
\begin{align}
	\frac{{\rm \partial} S}{{\rm \partial}V} = \frac{{\rm \partial}S(T, V)}{{\rm \partial}V} = T^{-1} \< P - P_0 \>.
	\label{p_dir_sol_temp}
\end{align}

\subsection{Calculation of the Hessian term \label{Hessian}}

The term $N^{-1} \log \det\hat{H}$ can be calculated by integrating over the crystal Brillouin zone. Hessian is a matrix of second derivatives with respect to displacement. 

First, let us define Hessian for the interaction of two atoms.
For this system, the Hessian is just a $3\times3$ matrix calculated as 
\[
H_{ij} = -\frac {\partial^2 E} {\partial r_i \partial r_j},\qquad i,j \in \{1,2,3\}.
\]
Here $r_i, r_j$ are the components of the vector $\bm{r}$, and $E$ is the potential energy of the system.
The Hessian matrix for a larger system is constructed by adding such $3 \times 3$ blocks of two-atom interactions.
For simplicity, we will only consider the $\mathbf{fcc}$ lattice case, which can be easily extended to an arbitrary lattice.

We start by denoting $\mathbf{fcc}$ the infinite fcc lattice with the point (0, 0, 0) excluded.
For a given vector in the k-space $\bm{k}$, cutoff $r_{cut}$ and the per-atom volume $V$, we define the Fourier transform of the Hessian matrix as
\[
\tilde{H}_{ij} = \sum_{\substack{\bm{r}\in\mathbf{fcc}\\ |\bm{r}|< r_{\rm cut}}} H_{ij}(\mathbf{r}) \big(1 - \cos(2\pi \bm{k}\cdot\bm{r})\big).
\]
We also define the auxiliary function
\[
\tilde{R}_{ij} = \sum_{\substack{\bm{r}\in\mathbf{fcc}\\ |\bm{r}|<1}} (1 - \cos(2\pi \bm{k}\cdot\bm{r})),
\]
which will help us to integrate the singularity at $\bm{k} = (0, 0, 0)$.

The vectors $\bm{k}$ lie in the Brillouin zone of fcc lattice.
The Brillouin zone $\mathbf{Br}$ of fcc lattice is the set of points $\bm{k}$ defined as:
\[
\mathbf{Br} \{\bm{k} : |\bm{k}|^2 \leq |\bm{k} - \bm{x}|^2, \bm{x}\in \mathbf{bcc}\}.
\]

Finally we express the value of $\log \det \hat{H}$ as follows:
\[
\log \det \hat{H} = 2 \left(\int\limits_{0 < k_1, k_2, k_3 < 1, \bm{k} \in \mathbf{Br}} \log \det (\tilde{H}/\tilde{R}) \right) + 6\left(\int\limits_{0 < k_1, k_2, k_3 < 1, \bm{k} \in \mathbf{Br}} \log\tilde{R}\right)
\]
(note that $\tilde{H}$ is a 3x3 matrix, hence the factor of $6$ in the second term).

The material-independent part, $\int_{0 < k_1, k_2, k_3 < 1, \mathbf{k} \in \mathbf{Br}} \log\tilde{R}$ is integrated beforehand with the accuracy of $10^{-15}$.

The error of this calculation, for the purpose of feeding it to the Gaussian process regression framework, is as the mean square of accuracy and difference of $\log \det \hat{H}$ values for the nearest cutoffs.

\subsection{Variance of a nonlinear functional \label{var}}

The melting point, defined as a system of equations \eqref{eq:trans_Y}, is a nonlinear functional of Gaussian processes.
In this case, we reduce the problem to a linear functional, obtained by expanding the original functional in the Taylor series around the mean of the Gaussian process.
We denote the original nonlinear functional as $\mathcal{F}=\mathcal{F}(S)$.
Then, the functional is linearized near the solution and the quantity of interest is now approximated with a linear functional, $\mathcal{F}(S) \approx \mathcal{F}(\overline{S}) + \<S-\overline{S}, J\>$,
where $\overline{S}$ is the mean predicted entropy, and $J$ is the gradient (Jacobian) of $\mathcal{F}$ evaluated at $\overline{S}$.
We can then evaluate variance, similarly to \eqref{eq:cov}
\begin{equation}
	\mathbb{V}[\mathcal F] \approx \mathbb{V}[J]
	=
	K(J, J) - K(J,\bm{X})^T K(\bm{X}, \bm{X})^{-1} K(J,\bm{X}).
	\label{eq:V}
\end{equation}

\end{document}